\pdfoutput=1

\documentclass[11pt,twoside,a4paper,cmspaper,final,collab]{cms-tdr}
\def\svnVersion{218721}\def\cmsCernNo{2013-015}\def\cmsCernDate{\today}\def\cmsMessage{Published in Physical Review D as \href{http://dx.doi.org/10.1103/PhysRevD.87.114015}{\doi{10.1103/PhysRevD.87.114015.}}}

\begin{document}\cmsNoteHeader{EXO-12-016}

\hyphenation{had-ron-i-za-tion}
\hyphenation{cal-or-i-me-ter}
\hyphenation{de-vices}

\RCS$Revision: 186276 $
\RCS$HeadURL: svn+ssh://svn.cern.ch/reps/tdr2/papers/EXO-12-016/trunk/EXO-12-016.tex $
\RCS$Id: EXO-12-016.tex 186276 2013-05-16 15:49:12Z alverson $
\newlength\cmsFigWidth
\ifthenelse{\boolean{cms@external}}{\setlength\cmsFigWidth{0.95\columnwidth}}{\setlength\cmsFigWidth{0.48\textwidth}}
\newlength\cmsFigWidthWide
\ifthenelse{\boolean{cms@external}}{\setlength\cmsFigWidthWide{0.95\columnwidth}}{\setlength\cmsFigWidthWide{0.75\textwidth}}
\ifthenelse{\boolean{cms@external}}{\providecommand{\cmsLeft}{top}}{\providecommand{\cmsLeft}{left}}
\ifthenelse{\boolean{cms@external}}{\providecommand{\cmsRight}{bottom}}{\providecommand{\cmsRight}{right}}
\cmsNoteHeader{EXO-11-094} % This is over-written in the CMS environment: useful as preprint no. for export versions
\title{Search for narrow resonances using the dijet mass spectrum in \texorpdfstring{$\Pp\Pp$ collisions at $\sqrt{s}=8$\TeV}{pp collisions at sqrt(s) = 8 TeV}}

\newcommand{\mjj}{\ensuremath{m_{jj}}\xspace}
\newcommand{\mjjFAT}{\ensuremath{m^\tiny{Wide}_{\text{jj}}}\xspace}
\newcommand{\mjjHLT}{\ensuremath{m^\tiny{HLT}_{\text{jj}}}\xspace}
\newcommand{\mjjRECO}{\ensuremath{m^\tiny{RECO}_{\text{jj}}}\xspace}
\newcommand{\mjjBIAS}{\ensuremath{m^\tiny{bias}_{\text{jj}}}\xspace}
\newcommand{\dphijj}{\ensuremath{\abs{\Delta\phi_{\text{jj}}}}\xspace}
\newcommand{\detajj}{\ensuremath{\abs{\Delta\eta_{jj}}}\xspace}
\newcommand{\dphijjFAT}{\ensuremath{\abs{\Delta\phi^\tiny{Wide}_{\text{jj}}}}\xspace}
\newcommand{\detajjFAT}{\ensuremath{\abs{\Delta\eta^\tiny {Wide}_{\text{jj}}}}\xspace}
\newcommand{\ptBIAS}{\ensuremath{\pt^\tiny{bias}}\xspace}
\newcommand{\ptRECO}{\ensuremath{\pt^\tiny{RECO}}\xspace}
\newcommand{\ptHLT}{\ensuremath{\pt^\tiny{HLT}}\xspace}
\newcommand{\Qstar}{\ensuremath{\mathrm{Q}^*}\xspace}
\newcommand{\minLimitDiquarkLowMass}{600\xspace}
\newcommand{\maxLimitDiquarkLowMass}{2600\xspace}
\newcommand{\minLimitWprimeLowMass}{600\xspace}
\newcommand{\maxLimitWprimeLowMass}{902\xspace}
\newcommand{\minLimitRSGLowMass}{600\xspace}
\newcommand{\maxLimitRSGLowMass}{646\xspace}

\newcommand{\lumiVal}{4.0\fbinv\xspace}
\newcommand{\lumiValUncert}{4.4\%\xspace}
\newcommand{\lumiValPhysicsDST}{0.13\fbinv\xspace}
\newcommand{\QCDMCscaleFactorDiWideJetMass}{1.34\xspace}
\newcommand{\chiSquare}{25.7\xspace}
\newcommand{\NDF}{32\xspace}
\newcommand{\jecUncert}{1.3\%\xspace}
\newcommand{\minMjjCut}{890\xspace}

\date{\today}

\abstract{ Results are presented of a search for the production of new
particles decaying to pairs of partons (quarks, antiquarks, or
gluons), in the dijet mass spectrum in proton-proton collisions at
$\sqrt{s} = 8$\TeV.  The data sample corresponds to an integrated
luminosity of 4.0\fbinv, collected with the CMS detector at the LHC in
2012.  No significant evidence for narrow resonance production is
observed.  Upper limits are set at the 95\% confidence level on the
production cross section of hypothetical new particles decaying to
quark-quark, quark-gluon, or gluon-gluon final states. These limits
are then translated into lower limits on the masses of new resonances
in specific scenarios of physics beyond the standard model.  The
limits reach up to 4.8\TeV, depending on the model, and extend
previous exclusions from similar searches performed at lower collision
energies.  For the first time mass limits are set for the
Randall--Sundrum graviton model in the dijet channel.}

\hypersetup{%
pdfauthor={CMS Collaboration},%
pdftitle={Search for narrow resonances using the dijet mass spectrum in pp collisions at sqrt(s) = 8 TeV},%
pdfsubject={CMS},%}
pdfkeywords={CMS, physics, search, exotica, dijet, resonance}}

\maketitle %maketitle comes after all the front information has been supplied

We report on a search for narrow dijet resonances in pp collisions at
$\sqrt{s}=8$\TeV. This search is applicable to for all new particles
for which the natural resonance width is small compared to the CMS
dijet mass resolution \cite{Harris:2011bh}.  The data sample
corresponds to an integrated luminosity of \lumiVal collected with the
Compact Muon Solenoid (CMS)~\cite{refCMS} at the CERN Large Hadron
Collider (LHC) in the spring of 2012.

Many extensions of the standard model (SM) predict the existence of
new massive objects that couple to quarks or antiquarks ($\cPq$ or
$\cPaq$) and gluons (\cPg), resulting in resonances in the dijet mass
spectrum. The most stringent bounds on these resonances come from
previous CMS~\cite{Khachatryan:2010jd, CMS:2012yf, EXO-11-095:2012pc}
and ATLAS~\cite{ATLAS2010,Aad:2011aj,Aad201237,ATLAS:2012pu}
searches. The results presented in this Letter extend the search
sensitivity to higher values of the resonance masses.

We consider the following specific models of narrow dijet resonances
produced via the s-channel: string
resonances~\cite{Anchordoqui:2008di,Cullen:2000ef}; ${E}_6$
diquarks~\cite{ref_diquark}; excited quarks assuming the dimensionless
constants accounting for possible deviations from the standard model
couplings to be $f=f'=fs=1$~\cite{ref_qstar,Baur:1989kv};
axigluons~\cite{ref_axi,Chivukula:2011ng}; color-octet
colorons~\cite{ref_coloron}; the S8 resonance predicted in technicolor
models~\cite{Han:2010rf}; new gauge bosons ($\PW^\prime$ and
$\cPZpr$)~\cite{ref_gauge}; Randall--Sundrum (RS) gravitons assuming
$k/\overline{M}_\text{Pl} = 0.1$, where $k$ is related to the
curvature of the 5th dimension and $\overline{M}_\text{Pl}$ is the
effective 4-D Planck scale ~\cite{ref_rsg}. More details on these
models and the parameters we assume can be found in
Ref.~\cite{Harris:2011bh, CMS:2012yf}.

A detailed description of the CMS experiment can be found
elsewhere~\cite{refCMS}.  The CMS coordinate system has the origin at
the centre of the detector.  The $z$ axis points along the directPrion
of the counter-clockwise beam;
$y$ is the vertical direction and $x$ is chosen to make a right-handed
coordinate system; $\phi$ is the azimuthal angle, $\theta$ is the
polar angle, and the pseudorapidity is defined as $\eta \equiv
-\ln[\tan(\theta/2)]$.  The central feature of the CMS apparatus is a
superconducting solenoid of 6\unit{m} internal diameter providing an
axial field of 3.8\unit{T}.  Within the field volume are located the
silicon pixel and strip tracker ($|\eta|<2.5$), as well as the barrel
and endcap calorimeters ($|\eta|<3$): a lead tungstate crystal
electromagnetic calorimeter and a brass/scintillator hadronic
calorimeter. An iron/quartz fiber calorimeter is located in the
forward region ($3<|\eta|<5$), outside the field volume. For
triggering purposes and to facilitate the reconstruction of
hadronizing particles as jets, the calorimeter cells are grouped into
towers projecting radially outward from the centre of the detector.

Offline particle candidates are reconstructed by using the particle
flow (PF) algorithm~\cite{PFPAS2009}, which categorizes the candidates
as muons, electrons (with associated bremsstrahlung photons), photons
(unconverted and converted), and charged/neutral hadrons. These PF
candidates are then clustered into jets using the anti-\kt
algorithm~\cite{antikt} with a distance parameter $R = 0.5$,
implemented in the \textsc{FastJet} package\cite{Cacciari:2005hq}. The
jet four-momentum, computed as the vectorial sum of the four-momenta
of the constituent PF candidates, is adjusted with corrections derived
from Monte Carlo (MC) simulations, test beam results, and pp collision
data~\cite{1748-0221-6-11-P11002}.  The corrections also account for
the presence of multiple pp collisions in the same or adjacent
bunch crossings (pileup interactions)~\cite{Cacciari:2007fd}.

Events are selected by requiring at least one reconstructed primary
vertex in each event within the range $|z|<24$\unit{cm}.  We select
jets with $\pt > 30\GeV$ and $|\eta|<2.5$ that meet identification
criteria based on the number of constituent particles and their energy
fractions~\cite{JME-10-003-PAS}. The other jets in the event are
ignored.  Events with fewer than two selected jets are discarded.

To improve the dijet invariant mass resolution, we account for final
state radiation (FSR) by forming a wide jet~\cite{Chatrchyan2011123,
CMS:2012yf} around each leading jet. The wide jets are formed by
clustering additional jets to the closest leading jet within a
distance $\Delta R = \sqrt{\Delta \eta^2 + \Delta \phi^2}< 1.1$.  The
four-momentum of each wide jet is computed as the sum of the
four-momenta of the constituent jets. To suppress background events
coming from quantum chromodynamics (QCD) processes, we require that
the pseudorapidity separation of the two wide jets satisfies
$\detajj<1.3$, and that both wide jets are reconstructed in the region
$|\eta|<2.5$. These angular requirements requirements maximize the
search sensitivity for isotropic decays of dijet resonances in the
presence of QCD background.  The dijet mass is given by
$\mjj=\sqrt{(E_{j1} + E_{j2})^2 - |\vec{p}_{j1} + \vec{p}_{j2}|^2}$,
where $E_{ji}$ and $\vec{p}_{ji}$ ($i=1,2$) are the energy and the
momentum of a wide jet.  We select events with $\mjj>\minMjjCut\GeV$
to maintain a fully efficient trigger as discussed below.

Events are filtered using a two-tier trigger system. Events satisfying
loose jet requirements at the first level (L1) are passed to the high
level trigger (HLT), where jets are clustered from PF candidates built
online.
Online jets with transverse momenta $\pt > 40\GeV$ and $|\eta|<3.0$
are used to compute $\HT$, the scalar sum of the jet $\pt$, and
$m_{jj}$, the invariant mass of the two wide jets. Events with
$\HT>650$\GeV or $m_{jj}>750$\GeV are accepted.  For the offline
analysis selection presented above, the combined L1 and HLT triggers
are found to be more than 99.9\% efficient.

\begin{figure}[htb]
  \begin{center}
    \includegraphics[width=\cmsFigWidthWide]{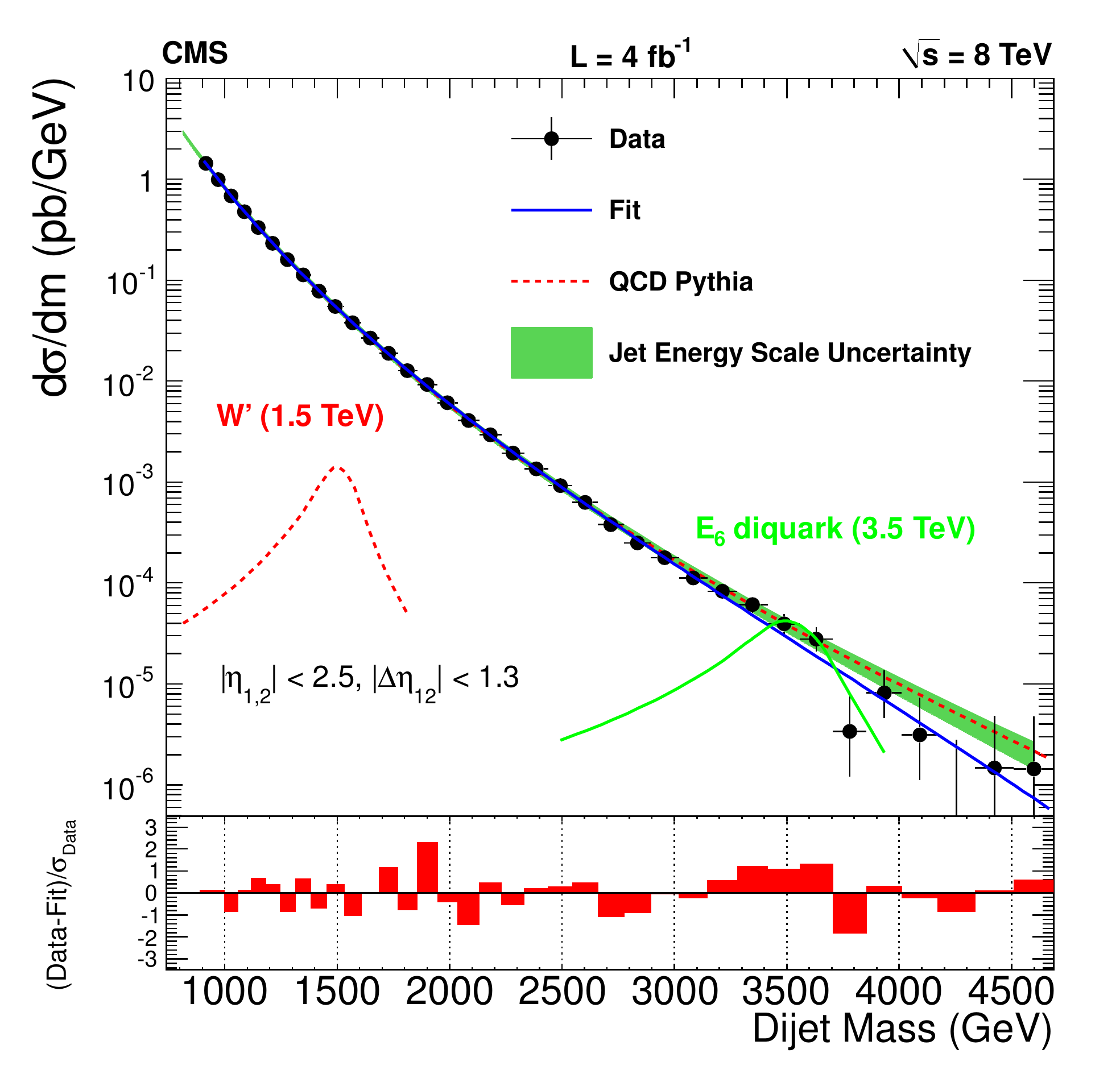}
   \caption{
The dijet mass spectrum from wide jets (points) compared with a smooth
fitted curve (solid line) and with the predicted QCD
background~\cite{Sjostrand:2006za} (dashed line).  The QCD background
curve has been normalized to the data (see text) and a linear
smoothing between the bins has been applied. The vertical bars on the
data points represent the statistical uncertainty, the horizontal bars
indicate the bin widths.  The shaded band shows the contribution from
the systematic uncertainty in the jet energy scale.  Also shown are
the predictions for a $\PW^\prime$ boson with a mass of 1.5\TeV, and an
${E}_6$ diquark with a mass of 3.5\TeV, obtained fitting the expected
distribution to a Crystal ball~\cite{Skwarnicki:1986xj} function and
normalizing the area to the predicted cross section.  The
bottom part of the plot displays the bin-by-bin residuals (data minus
the integral over a bin of the smooth function fitted to the data)
divided by the statistical uncertainty in the data.
    \label{figDataAndMC}}
  \end{center}
\end{figure}

We show in Fig.~\ref{figDataAndMC} the dijet mass distribution in bins
approximately equal in width to the dijet mass
resolution~\cite{Khachatryan:2010jd}. The data are compared with the
expected leading-order (LO) QCD background generated by using {\PYTHIA
v6.424}~\cite{Sjostrand:2006za}, including a
\GEANTfour-based~\cite{refGEANT} simulation of the CMS detector. This
approach follows closely that described in~\cite{bib_pythiatunes}, but
uses the CTEQ6L PDF (Z2 tune) instead of the CTEQ5L PDF (Z1 tune).
The QCD prediction uses a renormalization and factorization scale
$\mu=\pt$ of the hard-scattered partons and CTEQ6L1 parton
distribution functions~\cite{refCTEQ}, and has been normalized to the
data.  The normalization factor of ~\QCDMCscaleFactorDiWideJetMass was
found to be consistent with the next-to-leading order
K-factor~\cite{Ellis:1992en,Giele:1993dj}.  The shape of the {\PYTHIA}
prediction agrees with the data within the statistical precision.

For comparison we also display in Fig.~\ref{figDataAndMC} the shape
expected for a $\PW^\prime$ boson with a mass of 1.5\TeV and an ${E}_6$
diquark with a mass of 3.5\TeV. The signal samples are generated by
using {\PYTHIA} with the D6T tuning~\cite{Sjostrand:2006za} and the
same \GEANTfour-based CMS simulation used for the QCD background
sample. The predicted mass distributions have a Gaussian core from the
jet energy resolution and a tail towards lower mass values, primarily
due to FSR. The contribution of this low-mass tail to the line-shape
depends on the parton content of the resonance ($\cPq\cPq$,
$\cPq\cPaq$, $\cPq\cPg$ and $\cPg\cPg$). Resonances decaying to
gluons, which are more susceptible than quarks to the FSR, have a
larger tail. For high-mass resonances, there is also another
significant contribution depending on both parton distributions and
the natural width of the Breit--Wigner resonance shape: when the
resonance is produced by interaction of non-valence partons in the
proton, the low-mass component of the Breit--Wigner resonance shape is
amplified by a larger parton probability at low fractional momentum,
producing a large tail at low mass values.  The shapes shown for a
hypothetical $\PW^\prime$ boson and an ${E}_6$ diquark in
Fig.~\ref{figDataAndMC} result from Crystal
ball~\cite{Skwarnicki:1986xj} fits to the generated event
distributions.

The background from QCD multijet production is described by the
analytical function
\begin{equation*}
\frac{\rd\sigma}{\rd x} =
\frac{P_{0} (1 - x)^{P_{1}}}{x^{P_{2} + P_{3} \ln{(x)}}},
\end{equation*}
with the variable $x=\mjj/\sqrt{s}$ and four free parameters $P_0$,
$P_1$, $P_2$, and $P_3$.  This functional form has been used in
previous searches \cite{refCDFrun2,ATLAS2010,Khachatryan:2010jd,
Aad:2011aj} to describe both data and QCD predictions. The fit is
performed maximizing a binned likelihood, the bins being defined as in
Fig.~\ref{figDataAndMC}. The fit result, also shown in
Fig.~\ref{figDataAndMC}, has a chi-squared ($\chi^2$) of \chiSquare
for \NDF degrees of freedom. The bottom part of the figure shows the
difference between the data and the fit value, normalized to the
statistical uncertainty in the data. Assuming a pure $q \bar q$ final
state, the largest upward deviation of the data corresponds to a local
significance of 2.3 $\sigma$ and a global significance of 0.6 $\sigma$
once including the look elsewhere effect. Different assumptions on the
final state composition gives smaller values.

A data-driven determination of the background is obtained through a
smooth fit to the data. We use the dijet mass spectrum from wide jets,
the background parametrization, and the dijet resonance shapes to set
specific limits on new particles produced from and decaying to a same
parton pair $\cPq\cPq$ (or $\cPq\cPaq$), $\cPq\cPg$, and $\cPg\cPg$. A
separate limit is determined for each process (denoted simply
$\cPq\cPq$, $\cPq\cPg$, $\cPg\cPg$) because of the dependence of the
signal line-shape on the final state, induced by the different amount
of FSR for gluons and quarks.

The systematic uncertainty in the determination of the dijet mass is
dominated by the uncertainty in the jet energy
scale~\cite{1748-0221-6-11-P11002} and the uncertainty in the jet
energy resolution. The jet energy scale uncertainty translates into a
\jecUncert relative uncertainty in the dijet mass, roughly independent
from the mass value; it is propagated to the search by shifting the
reconstructed dijet mass of the signal by \jecUncert compared to the
nominal resolution value. The jet energy resolution uncertainty
translates into an uncertainty of 10\% in the resolution of the dijet
mass~\cite{1748-0221-6-11-P11002}; this uncertainty is propagated to
the search by smearing and unsmearing the reconstructed dijet mass of
the signal according to a Gaussian distribution with $\sigma$ fixed at
10\% of the mass value.

The precision of the overall signal normalization is limited by the
knowledge of the integrated luminosity
(\lumiValUncert)~\cite{LUMIPASSummer2012}. The statistical uncertainty
in the background parametrization introduces a systematic uncertainty
in the signal strength. We verified that the use of different
parametrizations for the description of the background has a
negligible effect compared to the statistical uncertainty in the data,
over the whole dijet spectrum.  Similarly, MC studies show that the
dependence of the signal mass shapes on the number of pileup
interactions is negligible.  The systematic uncertainties included in
this analysis reduce the lower limit on resonance masses by less than
${\sim}15$\GeV, depending on the model.

To set upper limits on the signal cross section we use a Bayesian
formalism~\cite{PDG2012-Bayes} with a uniform prior for the positive
signal cross section; a null probability is assigned to negative
values of the cross section; log-normal priors are used to model
systematic uncertainties, which are treated as nuisance parameters. We
calculate the posterior probability density as a function of resonance
cross section independently at each value of the resonance mass.  The
data are fitted to the background function plus a signal line-shape,
the signal cross section being a free parameter. The resulting fit
function with the signal cross section set to zero is used as the
background hypothesis. The uncertainty in the background shape is
incorporated by marginalizing over the background fit parameters (not
including the signal cross section) after diagonalizing the covariance
matrix to account for the correlations in the parameters.  This method
of using the data first to constrain the background fit and second to
extract the limit induces a bias in the coverage of the limits.  The
actual coverage was estimated for the $\cPq\cPq$ resonances to be
$92.1\pm0.4\%$, $95.2\pm0.4\%$ and $95.8\pm0.3\%$ at respective signal
masses of 1500, 2500 and 3000\GeV.

\begin{figure*}[tb]
  \begin{center}
    \includegraphics[width=\cmsFigWidth]{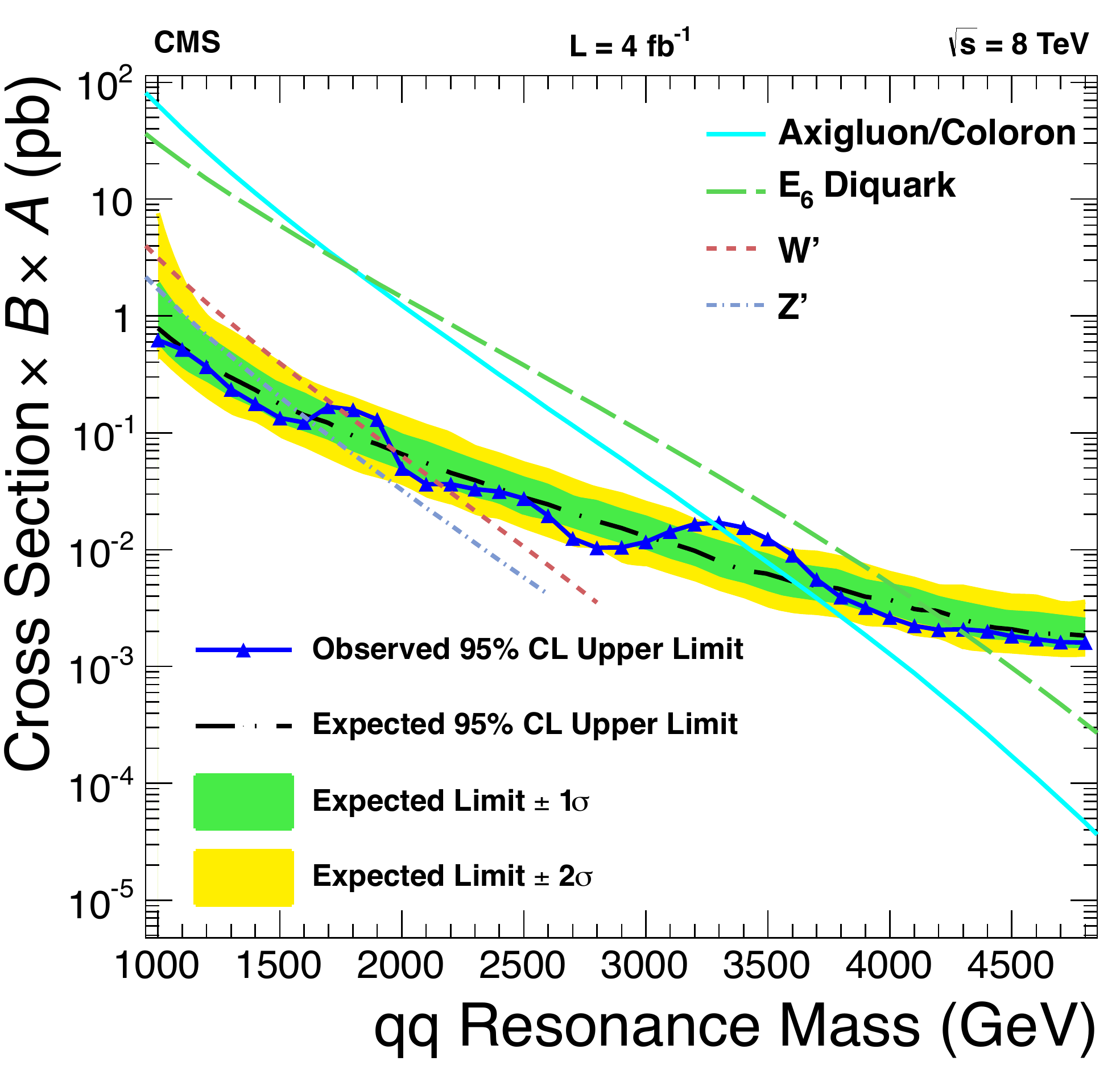}
    \includegraphics[width=\cmsFigWidth]{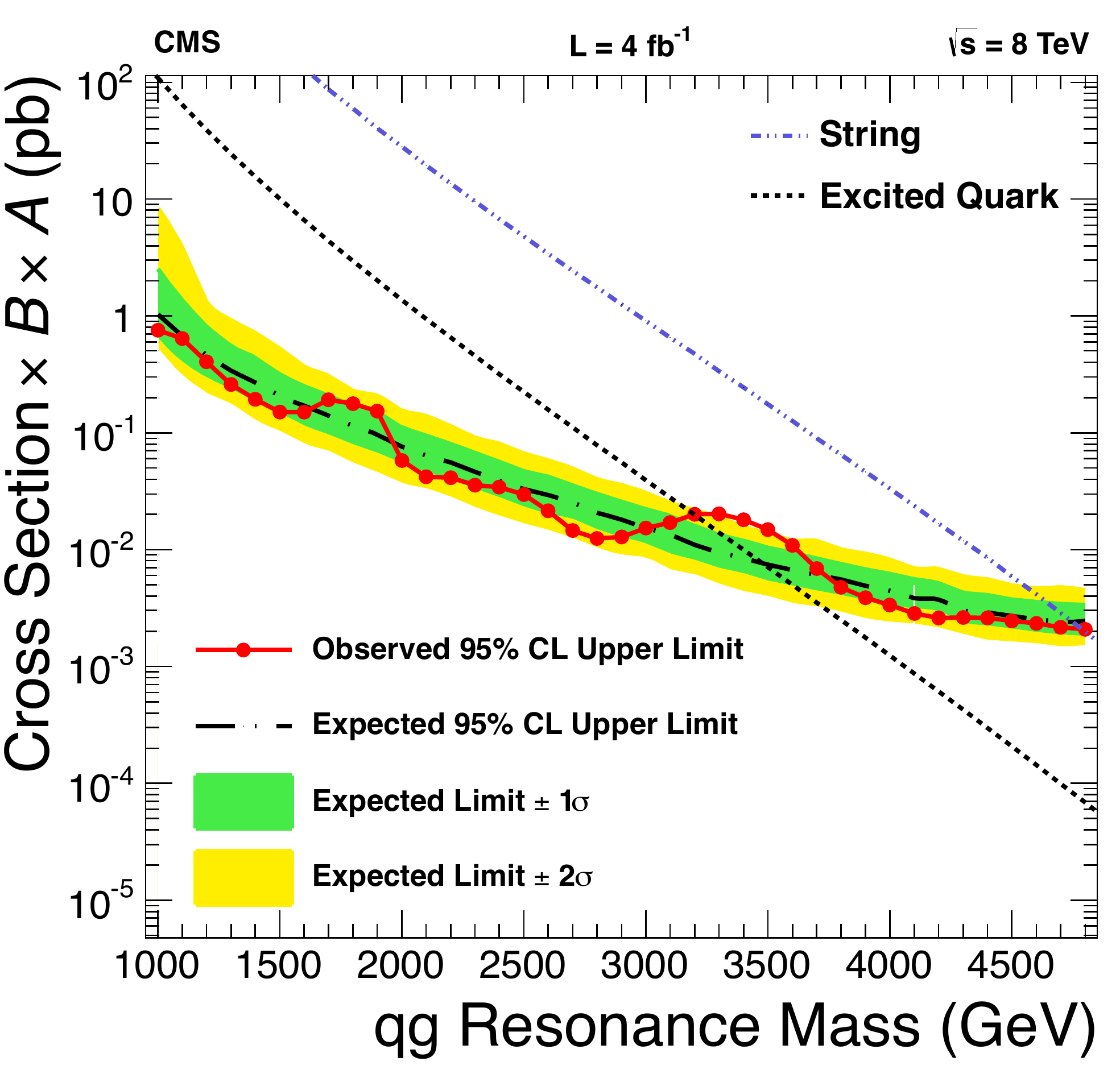}
    \includegraphics[width=\cmsFigWidth]{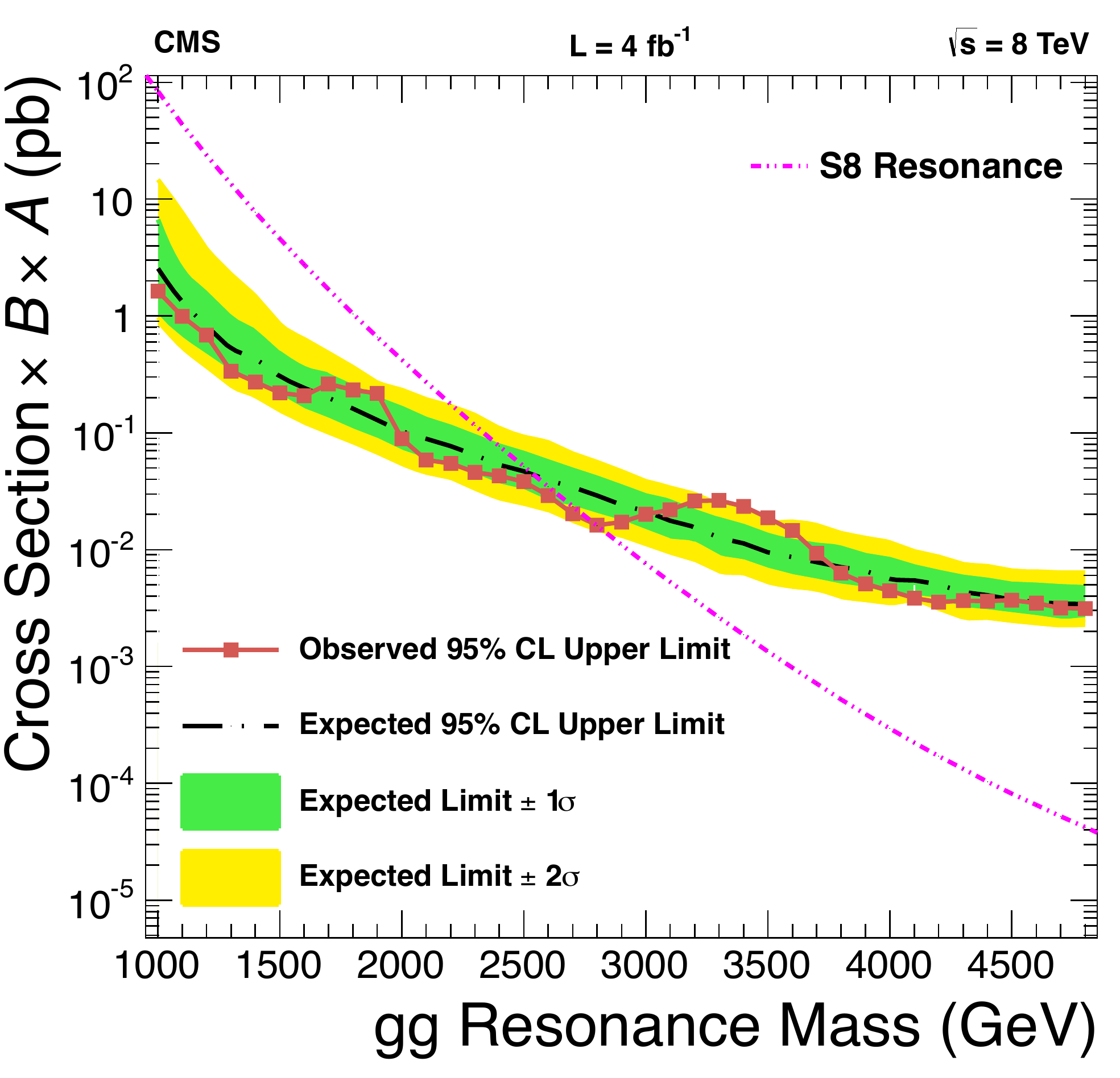}
    \includegraphics[width=\cmsFigWidth]{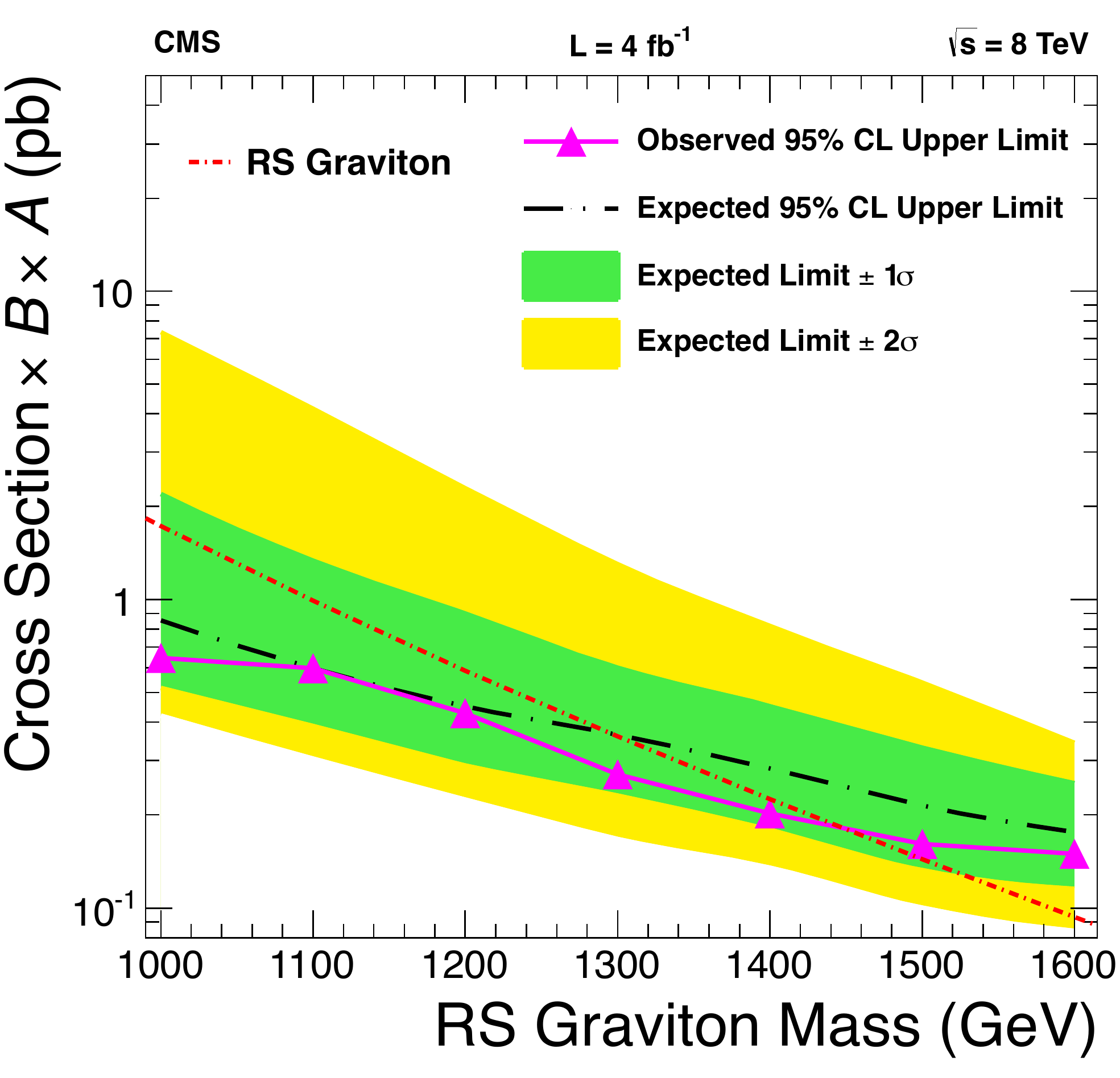}
    \caption{Observed upper limits at 95\% CL on $\sigma\times\mathcal{B}
    \times A$ for resonances decaying to $\cPq\cPq$, $\cPq\cPg$, and
    $\cPg\cPg$ final state (points and solid lines) compared with the
    expected limits (dot-dashed dark lines) and their variation at the
    1$\sigma$ and 2$\sigma$ levels (shaded bands).  Predicted cross
    sections for various resonance assumptions are shown. The signal
    shape for the RS model is obtained by weighting the shapes for
    $\cPq\cPq$ and $\cPg\cPg$ final states according to LO
    calculations of the relative branching
    fractions.\label{figExpectedLimit}}
  \end{center}
\end{figure*}

We show in Fig.~\ref{figExpectedLimit} the observed upper limits at
the 95\% confidence level (CL) on $\sigma\times B\times A$, \ie the
product of the cross section ($\sigma$), the branching fraction
($\mathcal{B}$) of the resonance into the relevant final state, and
the acceptance ($A$) for reconstructing two jets with $\detajj<1.3$
and $|\eta|<2.5$, for narrow resonances which decay into $\cPq\cPq$,
$\cPq\cPg$, and $\cPg\cPg$ final states.  For example the acceptance
for an isotropic decays is $A\approx 0.6$, roughly independent of
resonance mass.  For the RS graviton, which couples either to a pair
of gluons or to a quark-antiquark pair, the model-dependent limits on
cross section are derived using a weighted average of the $\cPq\cPq$
and $\cPg\cPg$ dijet mass shapes, where the weights correspond to the
relative branching fractions for these two final states, calculated at
LO~\cite{ref_rsg}.  The expected limits on cross sections shown in
Fig.~\ref{figExpectedLimit} are estimated with pseudo-experiments
generated using background shapes which are determined by
signal-plus-background fits to the data.

The observed and expected upper limits can be compared to the
predictions for $\sigma\times \mathcal{B}\times A$ before including
any detector simulation, in order to determine mass limits on new
particles. The calculations shown are obtained in the narrow-width
approximation using CTEQ6L1 parton distributions~\cite{refCTEQ}.  New
particles are excluded at the 95\% CL in mass regions for which the
theoretical curve lies above our upper limit for the appropriate final
state.

We list in Table~\ref{exclusionLimits_xsec} the observed cross section
limits for $\cPq\cPq$, $\cPq\cPg$, and $\cPg\cPg$ resonances, as a
function of the resonance mass. We determine the expected lower limit
on the mass of new resonances by comparing the expected cross section
limits to the model predictions. The observed and expected mass
exclusions are reported in Table~\ref{exclusionLimits_fat} for various
models.

\begin{table}[htbp]
  \centering
  \topcaption{Observed upper limits at the 95\% CL on
    $\sigma\times\mathcal{B} \times A$ for resonances decaying to
    $\cPq\cPq$, $\cPq\cPg$, and $\cPg\cPg$ final states as a function
    of the resonance mass. Experimental systematic uncertainties are
    taken into account in the limit
    calculation.\label{exclusionLimits_xsec}}
  \begin{scotch}{clll}
    Mass [\GeVns{}] & \multicolumn{3}{l}{Upper limit on $\sigma \times \mathcal{B} \times A$ (pb)} \\
     & \multicolumn{1}{l}{$\cPq\cPq$} &\multicolumn{1}{l}{$\cPq\cPg$} & \multicolumn{1}{l}{$\cPg\cPg$} \\
    \hline
    1000 & 0.62 & 0.75 & 1.6 \\
    1200 & 0.37 & 0.40 & 0.69 \\
    1400 & 0.18 & 0.19 & 0.27 \\
    1600 & 0.12 & 0.15 & 0.21 \\
    1800 & 0.16 & 0.18 & 0.23 \\
    2000 & 0.050 & 0.058 & 0.089 \\
    2200 & 0.036 & 0.041 & 0.055 \\
    2400 & 0.031 & 0.034 & 0.043 \\
    2600 & 0.019 & 0.021 & 0.029 \\
    2800 & 0.010 & 0.012 & 0.016 \\
    3000 & 0.012 & 0.015 & 0.020 \\
    3200 & 0.017 & 0.020 & 0.026 \\
    3400 & 0.016 & 0.018 & 0.024 \\
    3600 & 0.0090 & 0.011 & 0.015 \\
    3800 & 0.0039 & 0.0047 & 0.0063 \\
    4000 & 0.0026 & 0.0034 & 0.0045 \\
    4200 & 0.0021 & 0.0026 & 0.0036 \\
    4400 & 0.0020 & 0.0026 & 0.0036 \\
    4600 & 0.0017 & 0.0023 & 0.0035 \\
    4800 & 0.0016 & 0.0021 & 0.0031 \\
    \end{scotch}
\end{table}

\begin{table}[htbp]
  \centering
  \topcaption{Observed and expected exclusions at the 95\% CL on the
  mass of various resonances. Experimental systematic uncertainties
  are included.\label{exclusionLimits_fat}}
  \begin{scotch}{lccc}
    Model & Final & Observed & Expected\\
          & state  & excluded      &  excluded\\
          &        & mass range & mass range\\
          &        & [\TeVns{}] & [\TeVns{}] \\
    \hline
    String Resonance & $\cPq\cPg$  & [1.0, 4.78]  & [1.0, 4.75] \\
    Excited Quark & $\cPq\cPg$  & [1.0, 3.19]  & [1.0, 3.47] \\
    ${E}_6$ Diquark & $\cPq\cPq$  & [1.0, 4.28]  & [1.0, 4.16] \\
    Axigluon/Coloron & $\cPq\cPaq$  & [1.0, 3.27]  & [1.0, 3.60] \\
    S8 Resonance & $\cPg\cPg$  & [1.0, 2.79]  & [1.0, 2.54] \\
    $\PW^\prime$ Boson & $\cPq\cPaq$  & [1.0, 1.73]  & [1.0, 1.97] \\
    $\cPZpr$ Boson & $\cPq\cPaq$  & [1.0, 1.62]  & [1.0, 1.58] \\
    RS Graviton & $\cPq\cPaq+\cPg\cPg$  & [1.0, 1.45]  & [1.0, 1.29] \\
  \end{scotch}
\end{table}

In summary, a search for narrow resonances decaying into a pair of
jets has been performed using a data sample corresponding to an
integrated luminosity of \lumiVal collected in pp collisions at
$\sqrt{s}=8$\TeV. In the analyzed data sample there is no significant
evidence for new particle production.  Upper limits are set on the
product $\sigma\times \mathcal{B}\times A$ that can be applied to any
model of narrow dijet resonance production.  Specific lower limits are
presented on the mass of string resonances, excited quarks, axigluons,
colorons, S8 resonances, ${E}_{6}$ diquarks, $\PW^\prime$ and $\cPZpr$ bosons
starting at 1.0\TeV and extending up to 4.8\TeV for some models. In
most cases, these limits extend the previous exclusions obtained using
the dijet mass spectrum.  The first mass limit on Randall--Sundrum
gravitons decaying to dijets is shown.

We congratulate our colleagues in the CERN accelerator departments for
the excellent performance of the LHC and thank the technical and
administrative staffs at CERN and at other CMS institutes for their
contributions to the success of the CMS effort. In addition, we
gratefully acknowledge the computing centres and personnel of the
Worldwide LHC Computing Grid for delivering so effectively the
computing infrastructure essential to our analyses. Finally, we
acknowledge the enduring support for the construction and operation of
the LHC and the CMS detector provided by the following funding
agencies: BMWF and FWF (Austria); FNRS and FWO (Belgium); CNPq, CAPES,
FAPERJ, and FAPESP (Brazil); MEYS (Bulgaria); CERN; CAS, MoST, and
NSFC (China); COLCIENCIAS (Colombia); MSES (Croatia); RPF (Cyprus);
MoER, SF0690030s09 and ERDF (Estonia); Academy of Finland, MEC, and
HIP (Finland); CEA and CNRS/IN2P3 (France); BMBF, DFG, and HGF
(Germany); GSRT (Greece); OTKA and NKTH (Hungary); DAE and DST
(India); IPM (Iran); SFI (Ireland); INFN (Italy); NRF and WCU
(Republic of Korea); LAS (Lithuania); CINVESTAV, CONACYT, SEP, and
UASLP-FAI (Mexico); MSI (New Zealand); PAEC (Pakistan); MSHE and NSC
(Poland); FCT (Portugal); JINR (Armenia, Belarus, Georgia, Ukraine,
Uzbekistan); MON, RosAtom, RAS and RFBR (Russia); MSTD (Serbia); SEIDI
and CPAN (Spain); Swiss Funding Agencies (Switzerland); NSC (Taipei);
ThEPCenter, IPST and NSTDA (Thailand); TUBITAK and TAEK (Turkey); NASU
(Ukraine); STFC (United Kingdom); DOE and NSF (USA).

\bibliography{auto_generated}   % will be created by the tdr script.

\cleardoublepage \appendix\section{The CMS Collaboration \label{app:collab}}\begin{sloppypar}\hyphenpenalty=5000\widowpenalty=500\clubpenalty=5000\textbf{Yerevan Physics Institute,  Yerevan,  Armenia}\\*[0pt]
S.~Chatrchyan, V.~Khachatryan, A.M.~Sirunyan, A.~Tumasyan
\vskip\cmsinstskip
\textbf{Institut f\"{u}r Hochenergiephysik der OeAW,  Wien,  Austria}\\*[0pt]
W.~Adam, E.~Aguilo, T.~Bergauer, M.~Dragicevic, J.~Er\"{o}, C.~Fabjan\cmsAuthorMark{1}, M.~Friedl, R.~Fr\"{u}hwirth\cmsAuthorMark{1}, V.M.~Ghete, N.~H\"{o}rmann, J.~Hrubec, M.~Jeitler\cmsAuthorMark{1}, W.~Kiesenhofer, V.~Kn\"{u}nz, M.~Krammer\cmsAuthorMark{1}, I.~Kr\"{a}tschmer, D.~Liko, I.~Mikulec, M.~Pernicka$^{\textrm{\dag}}$, D.~Rabady\cmsAuthorMark{2}, B.~Rahbaran, C.~Rohringer, H.~Rohringer, R.~Sch\"{o}fbeck, J.~Strauss, A.~Taurok, W.~Waltenberger, C.-E.~Wulz\cmsAuthorMark{1}
% ...
\vskip\cmsinstskip
\dag:~Deceased\\
1:~~Also at Vienna University of Technology, Vienna, Austria\\
2:~~Also at CERN, European Organization for Nuclear Research, Geneva, Switzerland\\
% ...

\end{sloppypar}
\end{document}